\documentclass[pra,twocolumn]{revtex4}
\usepackage{graphicx}
\begin{document}

\bibliographystyle{unsrt}

\title{Cyclical Quantum Memory for Photonic Qubits}
\author{T.B. Pittman and J.D. Franson}
\affiliation{Johns Hopkins University,
Applied Physics Laboratory, Laurel, MD 20723}

\date{\today}

\begin{abstract}
We have performed a proof-of-principle experiment in which qubits encoded in the polarization states of single-photons from a parametric down-conversion source were coherently stored and read-out from a quantum memory device. The memory device utilized a simple free-space storage loop, providing a cyclical read-out that could be synchronized with the cycle time of a quantum computer.  The coherence of the photonic qubits was maintained during switching operations by using a high-speed polarizing Sagnac interferometer switch.
\end{abstract}

\maketitle

One of the most attractive features of an all-optical approach to quantum computing is that photons can serve as ideal carriers of quantum information in a modular system connecting various quantum logic gates and memory devices. However, the implementation of the memory devices and logic gates is difficult in an optical approach because photons are relatively difficult to store, and nonlinear interactions between single photons are typically very weak.  Although recent developments in linear optics quantum computing \cite{knill01a,franson02a} have shown that logic gates can be operated \cite{pittman02a} without direct nonlinear interactions, the implementation of a practical quantum memory for photons remains a challenging problem. In this paper we report on a proof-of-principle experimental demonstration of a such a quantum memory device.

Earlier suggestions for a quantum memory for photons have involved storing the quantum states in high-Q cavities \cite{maitre97} or reversibly transferring them to collective atomic excitations (see, for example \cite{fleischhauer00,kozhekin00}). In contrast, the quantum memory demonstrated here is based on the storage of single-photons in a simple free-space optical loop \cite{franson98}. The photons stored in this loop can be switched out after any number of round trips, providing a cyclical quantum memory (CQM) that could be synchronized with the cycle time of an optical quantum computer. 

The CQM presented here is an extension of our earlier work on a source of single-photons on pseudo-demand, which was realized by using straightforward 
polarization-based electro-optic switching techniques to release single-photons from a storage loop at a desired time \cite{pittman02c}. The key requirement in converting that system into the present cyclical quantum memory device was the development of interferometric-based switching techniques that work equally well for any polarization state, and are capable of maintaining the coherence of the polarization-encoded single-photon qubits. As will be described below, this was accomplished by using a high-speed electro-optic device to apply controlled $\pi$-phase shifts in a balanced polarizing Sagnac interferometer. During each passage through the interferometric switch, the output port taken by the photon could be controlled by the $\pi$-phase shifts, thereby sending the photon back into the storage loop for another cycle, or releasing it from the memory device.

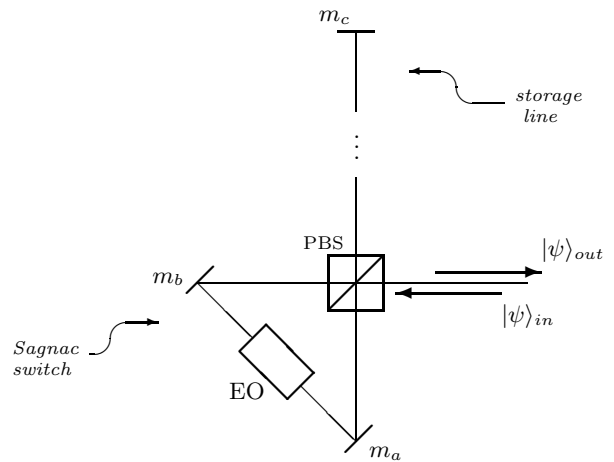
\begin{figure}[b]
\begin{center}
\begin{picture}(230,175)
\thicklines
\put(120,60){\framebox(20,20)}
\put(110,83){\scriptsize PBS}
\put(120,60){\line(1,1){20}}
\put(66,66){\line(1,1){10}}
\put(53,70){$m_{b}$}
\put(126,6){\line(1,1){10}}
\put(135,4){$m_{a}$}
\put(86,44){\line(1,-1){18}}
\put(86,44){\line(1,1){10}}
\put(104,26){\line(1,1){10}}
\put(96,54){\line(1,-1){18}}
\put(82,26){EO}
\put(185,66){\vector(-1,0){40}}
\put(185,55){$|\psi\rangle_{in}$}
\put(160,74){\vector(1,0){40}}
\put(200,80){$|\psi\rangle_{out}$}
\thinlines
\put(45,55){\vector(1,0){10}}
\put(45,49){\oval(16,12)[tl]}
\put(29,49){\oval(16,12)[br]}
\put(0,43){\scriptsize \em Sagnac}
\put(0,35){\scriptsize \em switch}
\put(195,70){\line(-1,0){125}}
\put(70,70){\line(1,-1){21}}
\put(109,31){\line(1,-1){21}}
\put(130,10){\line(0,1){100}}
\put(129,117){\vdots}
\put(130,135){\line(0,1){30}}
\thicklines
\put(123,165){\line(1,0){14}}
\put(116,169){$m_{c}$}
\thinlines
\put(160,150){\vector(-1,0){10}}
\put(160,144){\oval(16,12)[tr]}
\put(176,144){\oval(16,12)[bl]}
\put(176,138){\line(1,0){10}}
\put(190,139){\scriptsize \em storage}
\put(193,131){\scriptsize \em line}
\end{picture}
\end{center}
\vspace*{-.2in}
\caption{A schematic illustrating the basic features of the CQM, which utilizes a high-speed electro-optic polarizing Sagnac interferometer switch to maintain the coherence of stored polarization-encoded single-photon qubits.}
\label{fig:CQM} 
\end{figure}

A schematic illustrating the basic features of the CQM is shown in Figure \ref{fig:CQM}.  The polarizing Sagnac interferometer switch is formed by mirrors $m_{a}$ and $m_{b}$, and a polarizing beamsplitter (PBS) which reflects vertical polarizations and transmits horizontal polarizations  The electro-optic device (EO) inside the interferometer is a Pockels cell that is configured in such a way that when it is turned ``on'' it will rotate horizontal polarizations into vertical, and vice-versa ($|H\rangle\leftrightarrow|V\rangle$), and when it is turned ``off'' it will do nothing.  The storage line is formed by the upper port of the PBS and mirror $m_{c}$.

The logical values 0 and 1 are represented by the horizontal and vertical polarization states of a single-photon, so that an input qubit has the form $|\psi\rangle_{in}=\alpha|H\rangle+\beta|V\rangle$, where $\alpha$ and $\beta$ are arbitrary coefficients.  A qubit enters the CQM device from the right, and much of the operation of the device can be understood by first considering the case when the Pockels cell is never turned ``on''.  In this case, the vertical component of the incident qubit travels clockwise through polarizing Sagnac and is reflected up into the storage line, while the horizontal component travels counter-clockwise and is transmitted up into the storage line.  Upon reflection from mirror $m_{c}$, the process essentially runs in reverse, providing an output state $|\psi\rangle_{out}$ after one cycle of the quantum memory.

If there were no losses and no polarization-dependent phase shifts, the output state would emerge in the same coherent superposition state as the input, $|\psi\rangle_{out}=|\psi\rangle_{in}$.  Although there were, of course, losses and small birefringent phase-shifts in our experiment, the practical benefit of the polarizing Sagnac interferometer is that the counter-propagating horizontal and vertical components experience the same physical path, essentially eliminating the much larger thermally or vibrationally induced phase shifts that would otherwise ruin the coherence of the qubit in, for example, a comparable Mach-Zehnder interferometer.

In order for the CQM to store the qubit for more than one cycle, the Pockels cell must be quickly turned ``on'' while the photon is propagating in the storage line for the first time.  Upon subsequent passes through the Sagnac interferometer, the counter-propagating horizontal and vertical components of the qubit are therefore repeatedly flipped ($|H\rangle\leftrightarrow|V\rangle$), and the photon remains trapped in the device as long as the Pockels cell is left ``on''.  When the Pockels is finally turned ``off'', it can be seen that the final values of the counter-propagating  components are those required to release the photon from the device.

Consequently, one of the interesting features of the CQM is that qubits stored for an even number of cycles emerge as the bit-flipped value of input qubit,  $|\psi\rangle_{even}=\hat{\sigma}_{x}|\psi\rangle_{in}=\alpha|V\rangle+\beta|H\rangle$, while qubits released after an odd number of cycles are not bit-flipped, 
$|\psi\rangle_{odd}=|\psi\rangle_{in}$.
For quantum computing applications, the feed-forward control techniques that we have previously demonstrated 
\cite{pittman02b} could be used to re-flip qubits read-out after even numbers of cycles in the CQM; alternatively, the CQM round-trip time could be designed to be half the cycle time of the computer.

\begin{figure}[t]
\begin{center}
\begin{picture}(140,140)
\put(0,20){\line(1,0){120}}
\put(20,0){\line(0,1){120}}
\put(125,18){$|H\rangle$}
\put(16,127){$|V\rangle$}
\thicklines
\put(20,20){\vector(2,1){80}}
\put(20,20){\vector(1,2){40}}
\put(105,60){$|\psi\rangle_{in}$}
\put(55,107){$\hat{\sigma}_{x} |\psi\rangle_{in}$}
\put(40,22){$30^{o}$}
\thinlines
\put(60,60){\oval(50,50)[tr]}
\put(60,85){\vector(-1,0){2}}
\put(67,72){\scriptsize \em bit}
\put(66,64){\scriptsize \em flip}
\end{picture}
\end{center}
\vspace*{-.25in}
\caption{A graphical representation of the predicted output states of the CQM for an arbitrarily chosen state of the input qubit, 
$|\psi\rangle_{in}= \frac{\sqrt{3}}{2}|H\rangle + \frac{1}{2}|V\rangle$, which corresponds to a linear polarization state at $30^{o}$.  The bit-flipped value corresponds to a linear polarization state at $60^{o}$.}
\label{fig:output} 
\end{figure}
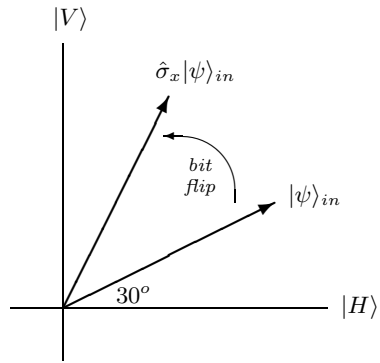

For all of the data that will be presented in this paper, the qubits were prepared in the arbitrarily chosen superposition state:

\begin{equation}
|\psi\rangle_{in} = \frac{\sqrt{3}}{2}|H\rangle + \frac{1}{2}|V\rangle
\label{eq:psiin}
\end{equation}

\noindent which corresponds to a linear polarization state at $30^{o}$. A graphical representation of the expected output qubits after both even and odd numbers of cycles is shown in Figure \ref{fig:output}.

 A more important consequence of the repeated flipping of the polarization components is that certain types of errors automatically correct themselves.  For example, unwanted polarization-dependent phase shifts imparted in the storage line affect each of the two components of the original qubit an equal number of times for those qubits switched out after an even number of round trips.  As a result, the accumulated relative phase-shift becomes an overall phase-shift that simply factors out of the final state.  Similarly, the net relative phase shift error of this kind for qubits switched out after an odd number $n$ of cycles is only that due to the final round-trip, and not the accumulation of the phase errors due to the previous $n-1$ round-trips. 

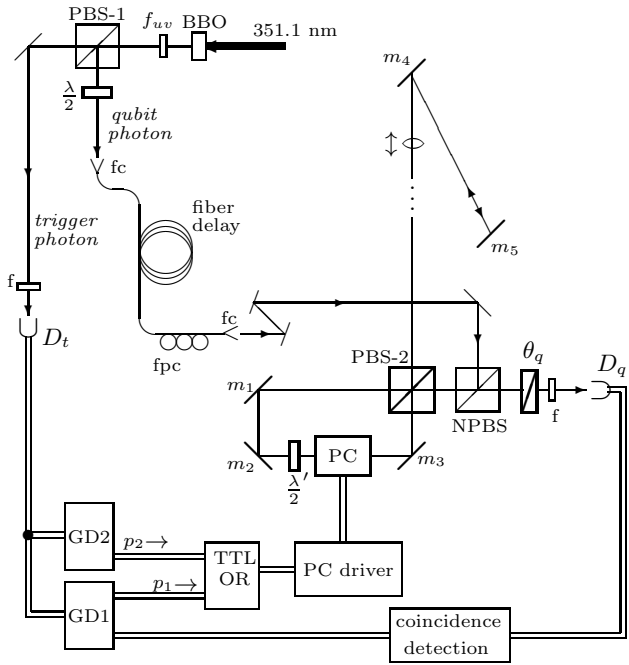
\begin{figure}[t]
\begin{center}
\begin{picture}(260,260)
\put(105,247){\vector(-1,0){30}}
\put(105,248){\vector(-1,0){30}}
\put(105,249){\vector(-1,0){30}}
\put(93,252){\scriptsize 351.1 nm}
\put(70,243){\framebox(5,10)}
\put(66,255){\scriptsize BBO}
\put(70,248){\line(-1,0){10}}
\put(58,244){\framebox(2,8)}
\put(51,256){\scriptsize $f_{uv}$}
\put(23,258){\scriptsize PBS-1}
\put(26,240){\framebox(16,16)}
\put(26,240){\line(1,1){16}}
\put(34,248){\line(0,-1){15}}
\thicklines
\put(29,229){\framebox(10,3)}
\thinlines
\put(20,227){$\frac{\lambda}{2}$}
\put(39,220){\scriptsize \em qubit}
\put(38,213){\scriptsize \em photon}
\put(34,229){\vector(0,-1){23}}
\put(58,248){\line(-1,0){50}}
\put(3,243){\line(1,1){10}}
\put(8,248){\vector(0,-1){50}}
\put(8,233){\line(0,-1){75}}
\put(10,180){\scriptsize \em  trigger}
\put(10,173){\scriptsize \em  photon}
\put(4,156){\framebox(8,2)}
\put(0,157){\scriptsize f}
\put(8,156){\vector(0,-1){10}}
\thicklines
\put(130,128){\scriptsize PBS-2}
\put(145,110){\framebox(16,16)}
\put(145,110){\line(1,1){16}}
\put(107,88){\framebox(3,10)}
\put(105,78){$\frac{\lambda}{2}^{\prime}$}
\put(117,86){\framebox(20,14){\scriptsize PC}}
\put(195,110){\framebox(5,15)}
\put(195,110){\line(1,3){5}}
\put(195,130){$\theta_{q}$}
\thinlines
\put(168,102){\scriptsize NPBS}
\put(170,110){\framebox(16,16)}
\put(170,110){\line(1,1){16}}
\put(200,118){\line(1,0){5}}
\put(205,114){\framebox(2,8)}
\put(206,106){\scriptsize f}
\put(207,118){\vector(1,0){12}}
\put(195,118){\line(-1,0){100}}
\put(95,118){\line(0,-1){25}}
\put(95,93){\line(1,0){12}}
\put(110,93){\line(1,0){7}}
\put(137,93){\line(1,0){16}}
\put(153,93){\line(0,1){90}}
\put(152,186){\vdots}
\put(153,198){\line(0,1){40}}
\put(153,237){\line(1,-2){30}}
\put(177,189){\vector(1,-2){3}}
\put(177,189){\vector(-1,2){3}}
\put(149,210){$\frown$}
\put(149,208){$\smile$}
\put(143,209){$\updownarrow$}
\thicklines
\put(148,233){\line(1,1){10}}
\put(142,242){\scriptsize $m_{4}$}
\put(178,172){\line(1,1){10}}
\put(183,170){\scriptsize $m_{5}$}
\put(148,88){\line(1,1){10}}
\put(155,90){\scriptsize $m_{3}$}
\put(90,98){\line(1,-1){10}}
\put(83,88){\scriptsize $m_{2}$}
\put(90,113){\line(1,1){10}}
\put(83,118){\scriptsize $m_{1}$}
\thinlines
\put(31,200){$\vee$}
\put(39,200){\scriptsize fc}
\put(42,200){\oval(16,12)[bl]}
\put(42,188){\oval(16,12)[tr]}
\put(50,188){\line(0,-1){43}}
\put(58,145){\oval(16,12)[bl]}
\put(57,139){\line(1,0){25}}
\put(81,137){$<$}
\put(81,143){\scriptsize fc}
\put(60,168){\circle{18}}
\put(60,170){\circle{18}}
\put(60,172){\circle{18}}
\put(70,185){\scriptsize fiber}
\put(70,178){\scriptsize delay}
\put(61,136){\circle{6}}
\put(67,136){\circle{6}}
\put(73,136){\circle{6}}
\put(55,125){\scriptsize fpc}
\put(88,139){\vector(1,0){12}}
\put(100,139){\line(1,0){5}}
\put(105,139){\line(-1,1){12}}
\put(93,151){\vector(1,0){35}}
\put(120,151){\line(1,0){58}}
\put(178,151){\vector(0,-1){16}}
\put(178,135){\line(0,-1){17}}
\put(173,156){\line(1,-1){10}}
\put(103,137){/}
\put(91,149){/}
\put(4,138){\large$\cup$}
\put(13,135){$D_{t}$}
\put(7,138){\line(0,-1){106}}
\put(9,138){\line(0,-1){104}}
\put(7,32){\line(1,0){15}}
\put(9,34){\line(1,0){13}}
\put(7,62){\line(1,0){15}}
\put(9,64){\line(1,0){13}}
\put(8,63){\circle*{4}}
\put(109,40){\framebox(40,20){\scriptsize PC driver}}
\put(126,60){\line(0,1){26}}
\put(128,60){\line(0,1){26}}
\put(75,35){\framebox(20,25)}
\put(78,52){\scriptsize TTL}
\put(80,44){\scriptsize OR}
\put(95,51){\line(1,0){14}}
\put(95,49){\line(1,0){14}}
\put(75,56){\line(-1,0){35}}
\put(75,54){\line(-1,0){35}}
\put(44,59){\scriptsize $p_{2}$}
\put(52,58){$\rightarrow$}
\put(75,41){\line(-1,0){35}}
\put(75,39){\line(-1,0){35}}
\put(55,43){\scriptsize $p_{1}$}
\put(63,42){$\rightarrow$}
\put(22,50){\framebox(18,25){\scriptsize GD2}}
\put(22,20){\framebox(18,25){\scriptsize GD1}}
\put(40,26){\line(1,0){105}}
\put(40,24){\line(1,0){105}}
\put(145,15){\framebox(45,20)}
\put(147,28){\scriptsize coincidence}
\put(150,18){\scriptsize detection}
\put(220,115){\large$\supset$}
\put(223,125){$D_{q}$}
\put(227,119){\line(1,0){7}}
\put(227,117){\line(1,0){5}}
\put(234,119){\line(0,-1){95}}
\put(232,117){\line(0,-1){91}}
\put(234,24){\line(-1,0){44}}
\put(232,26){\line(-1,0){42}}
\end{picture}
\end{center}
\vspace*{-.3in}
\caption{A schematic of the experimental set-up used for a laboratory demonstration the CQM. Details and symbols are described in the text.}
\label{fig:experiment} 
\end{figure}

A schematic representation of the actual experiment, which utilized true heralded single-photons from a parametric down-conversion source \cite{hong86}, is shown in Figure \ref{fig:experiment}. As shown in the upper part of Figure \ref{fig:experiment}, our down-conversion source consisted of a 1 mm thick BBO crystal pumped by roughly 30 mW of the 351.1 nm line of a continuous-wave Argon-Ion laser. The crystal was cut for Type-II degenerate phase matching and produced pairs of collinearly propagating, but orthogonally polarized photons at 702.2 nm \cite{rubin94}. The photons of a given pair were separated by a polarizing beamsplitter (PBS-1). The detection of the ``trigger photon'' by detector $D_{t}$ signalled the presence of the ``qubit photon'', and provided a relative start time for the cyclical device.  

The polarizing Sagnac interferometer switch, which can be seen in roughly the center of Figure \ref{fig:experiment}, was formed by polarizing beamsplitter PBS-2, and mirrors $m_{1}$ through $m_{3}$. The Pockels Cell (PC) was placed inside the Sagnac, with its fast axis rotated by $45^{o}$ from the vertical direction. An additional half-wave plate (labelled $\frac{\lambda}{2}^{\prime}$) was inserted next to the PC; this half-wave plate was usually oriented so that it had no effect, but it could be rotated as needed for various diagnostic tests during initial alignment.

The storage line was formed by the upper port of PBS-2 and mirrors $m_{4}$ and $m_{5}$ (the storage line was folded due to space constraints). The length of the storage line was roughly 3.5 m which, when combined with the .5 m length of the Sagnac, gave a total round-trip time of roughly 13.3 ns, which was longer than the 10 ns rise-time of the Pockels cell, as required \cite{pittman02c}. In order to minimize beam divergence and maximize the possible storage time (number of cycles) of the CQM, a 1 m focal length lens was placed in the storage line, and longitudinally adjusted to form a standard 2f Gaussian transmission line.

The qubit state preparation was implemented by using a half-wave plate (labelled $\frac{\lambda}{2}$) to rotate the vertical polarization of the qubit photon to any desired linear polarization state. 
The state of the output qubit was measured using a polarization analyzer ($\theta_{q}$) and another single-photon detector $D_{q}$. Background noise was reduced to a negligible level by using 10 nm bandwidth filters (f) centered at 700 nm in front of the trigger and qubit photon detectors.

In order to allow for the time required to produce and process the classical detection signal of the trigger photon, the qubit photon was delayed by a 120 m fiber optic delay line before being sent into the CQM \cite{pittman02c}. A standard fiber polarization controller (fpc) was used to negate the effects of birefringence in the fiber, as well as to pre-compensate for small birefringent phase-shifts of three additional steering mirrors (not labelled) and a 50/50 non-polarizing beamplitter (NPBS) that was used to couple the qubits into the CQM input channel.

The electronics used for the real-time active switching, which are seen in the lower part of Figure \ref{fig:experiment}, are described in detail in reference \cite{pittman02c}.  Briefly, the ``on'' to ``off'' transitions of the Pockels Cell were driven by the front and back edges of two different pulses ($p_{1}$ and $p_{2}$ from delay generators GD1 and GD2), which allowed the use of TTL pulses whose width (100ns) was greater than the cycle time. 

The accumulation of data involved the use of conventional coincidence counting techniques to record the time of arrival of the qubit photon at detector $D_{q}$ relative to the arrival time of the trigger photon at detector $D_{t}$.  Accumulated histograms of data would therefore show various peaks separated by 13.3 ns time intervals, corresponding to different numbers of round trips through the CQM.

\begin{figure}[b]
\vspace*{-.25in}
\hspace*{-.1in}
\includegraphics[width=3.5in]{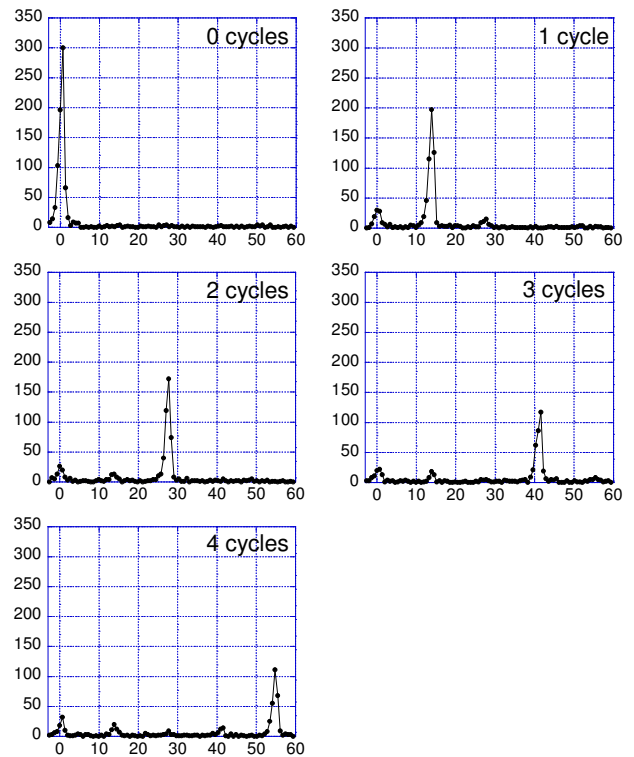}
\vspace*{-.6in}
\caption{A demonstration of qubit storage and active switch-out after a chosen number of cycles in the CQM. Each of the plots shows the number of coincidence counts per 600 sec. (vertical axes) as a function of the time of arrival in ns (horizontal axes) of the qubit photon at detector $D_{q}$ relative to the detection of the trigger photon at detector $D_{t}$.}
\label{fig:peaks}
\end{figure}

The data shown in Figure \ref{fig:peaks} demonstrates the ability to actively switch out the qubit after a chosen number of cycles through the CQM.
 For the data shown in Figure \ref{fig:peaks}, the qubit polarization analyzer $\theta_{q}$ was removed in order to clearly demonstrate the switching performance and assess any losses in the CQM. The area under the peak of interest in each successive plot is seen to decrease in approximate agreement with the 19\% loss per cycle that was measured in auxiliary experiments \cite{loss}. Of this 19\%, we estimate that roughly 15\% was due to reflection and transmission losses of the various optical components, with the remaining 4\% being due to imperfect focussing in the 2f Gaussian transmission line. Small switching errors \cite{pittman02c} which resulted in undesired noise peaks can also be seen in Figure \ref{fig:peaks}.

\begin{figure}[t]
\hspace*{-.1in}
\includegraphics[width=3.5in]{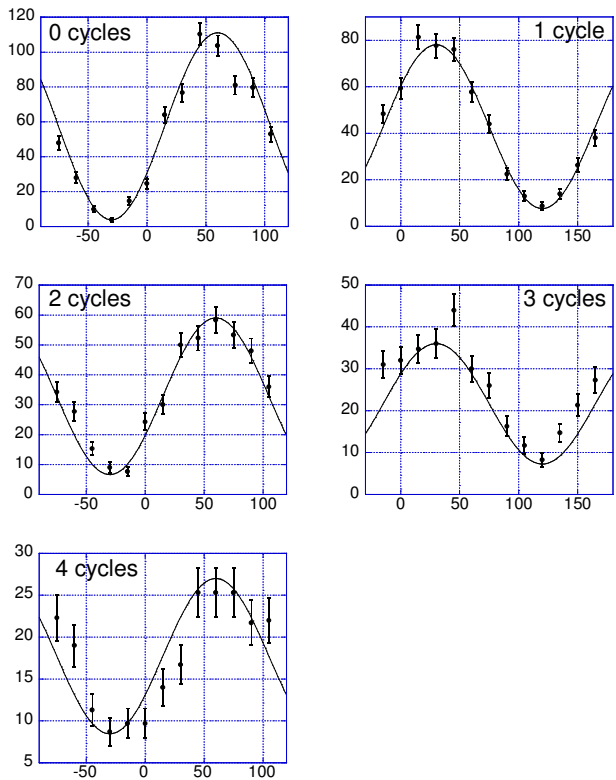}
\vspace*{-.5in}
\caption{Test of the coherence of the CQM device. The data shows a plot of the number of coincidence counts per 120 sec. (vertical axes) as a function of the qubit polarization analyzer $\theta_{q}$ setting in degrees (horizontal axes) for each of the five peaks shown in Figure \protect\ref{fig:peaks}. The solid lines (Cosine-squared curves centered at the expected output angle) are simply meant as guides to the eye.  The input qubits were linearly polarized at $30^{o}$, and the expected bit-flipping of those qubits stored for an even number of cycles is clearly seen.}
\label{fig:coherence}
\end{figure}

A test of the key feature of the CQM, the ability to maintain the coherence of the stored qubits, is summarized in Figure \ref{fig:coherence}.   For the data shown in Figure \ref{fig:coherence}, the qubit polarization analyzer $\theta_{q}$ was used to measure the polarization states of the stored qubits and compare them with the predicted output states shown in Figure \ref{fig:output}.  For qubits stored for an odd number of cycles the output states in Figure \ref{fig:coherence} are seen to be in good agreement with the linear input polarization state of $30^{o}$, while for even number cycles the expected (bit-flipped) linear polarization output state at $60^{o}$ is evident.

For each of the successive plots in Figure \ref{fig:coherence} the maximum counting rate and the quality of the output state is seen to decrease for qubits stored for longer times, as might have been expected. We believe the degradation of the qubit quality was primarily due to small misalignments of the mirrors, which produce spatial offsets that accumulate with each round-trip and can reduce the spatial overlap of the horizontal and vertical components in the output, thereby reducing the interference conditions required to maintain the coherence of the qubit.  We expect to be able to greatly reduce these errors by using fiber optic components in future implementations of the CQM.

In conclusion, we have performed a proof-of-principle demonstration of a new type of all-optical cyclical quantum memory device (CQM) based on the storage of photonic qubits in a simple free-space optical loop.  The ability to maintain the coherence of the qubits was accomplished here by the application of electro-optic-based controlled $\pi$-phase shifts in a balanced polarizing Sagnac interferometer switch \cite{mz}. 
An optical approach to quantum computing will most likely involve the use of trains of intense laser pulses, such as those from mode-locked Ti-Sapphire lasers.  Because these pulse trains provide a natural clock cycle, a cyclical quantum memory device of the kind presented here should be ideally suited for an optical approach to quantum computing, and these initial experiments may provide a first step in that direction.

This work was supported in part by ONR, ARO, NSA, ARDA, and IRAD.  We would like to acknowledge useful discussions with B.C. Jacobs, M.J. Fitch, and M.M. Donegan.




\begin{thebibliography}{50}
\bibitem{loss} A detailed description of the storage-loop loss measurement techniques can be found in reference \protect\cite{pittman02c}.
\bibitem{mz} For applications in which the output qubit is required to be in a different mode than the input, the polarizing Sagnac interferometer switch presented here could be generalized into, for example, an analogous polarizing Mach-Zehnder interferometer switch formed by two polarizing beamsplitters with Pockels Cells in each arm.
\end{thebibliography}
\end{document}